\documentclass[osajnl,two column,superscriptaddress,10pt]{revtex4-1} 
\usepackage{amsmath,amssymb,graphicx}
\begin{document}

\title{Self-focused beams to couple light into a whispering-gallery mode resonator }

\email{Corresponding author: kphanhuy@univ-fcomte.fr}
\author{Kien Phan~Huy}
\affiliation{FEMTO-ST Institute, UMR CNRS 6174, Universit\'e de Franche-Comt\'e, 16 Route de Gray,
25000 Besan\c con, France}

\author{Jassem Safioui}
\affiliation{FEMTO-ST Institute, UMR CNRS 6174, Universit\'e de Franche-Comt\'e, 16 Route de Gray,
25000 Besan\c con, France}

\author{Jean-Yves Rauch}
\affiliation{FEMTO-ST Institute, UMR CNRS 6174, Universit\'e de Franche-Comt\'e, 16 Route de Gray,
25000 Besan\c con, France}

\author{Patrice F\'eron}
\affiliation{FOTON-Syst\`emes Photoniques (CNRS-UMR 6082), ENSSAT, 6 rue de Kerampont, CS 80518, 22305 Lannion cedex, France}

\author{Mathieu Chauvet}
\affiliation{FEMTO-ST Institute, UMR CNRS 6174, Universit\'e de Franche-Comt\'e, 16 Route de Gray,
25000 Besan\c con, France}


\begin{abstract}
We propose an original method to couple light into a whispering-gallery mode resonator. This method benefits from the mode selectivity and robustness of the prism-coupling along with the single-mode propagation of the fiber taper. It consists in a prism shaped crystal with a waveguide inscribed inside it. The waveguide is self-inscribed in-situ by beam self-trapping to allow an optimum coupling to a given resonator. 


\end{abstract}


\maketitle 


Since the seminal paper of Tien and Ulrich \cite{CouplagePrism}, prism coupling has been a very successful coupling method. It was the natural choice in the late 80's to characterize high-Q resonators \cite{HighQSoviet}. Later on, a new method involving side-polished or fused-fiber taper took over with the benefit of optical fiber compatibility \cite{TaperPolished,TaperKnight}. In the prism-coupling method, the total internal reflexion (TIR) upon the internal face of a prism is used to produce an evanescent field that couples with a whispering-gallery mode (WGM). A proper adjustment of the angle of incidence enables to fine tune the propagation constant and insure the phase matching with the WGM. For best coupling, the incident-beam size should be as close as possible to the WGM size. To satisfy this criterion, light is usually focused upon the coupling surface. However, focusing too much eventually leads to the coupling of neighbour WGMs with close propagating constants. Consequently, a compromise had to be found leading to an awkward method. In the tapered fiber method, a standard optical fiber is pulled when heated by a gas-burner or an electric heater. The waist of the fiber is then adiabatically reduced along the fiber. As the guided mode propagates, it reaches the smallest diameter region (typically 1~$\mu$m) where the light is guided by TIR at the interface between silica and air. Due to the small core diameter, the evanescent  field extents in air, thus couples to the WGM resonator. Unfortunately, this method suffers from the inherent fragility of the device. All the more, a good control of the fusion-pulling process is needed to reach the proper propagating constant, hence, a less flexible method compared to the prism coupling \cite{TaperKnight}. 

In this paper, we propose an alternative method combining the flexibility of the prism-coupling and a guided propagation. We use a prism-coupling set-up that involves a prism made of a nonlinear material. The prism-coupling configuration enables to fine tune the phase matching at low power. At high power, the light beam induces its own waveguide thanks to the optical nonlinearity and is thus self-confined, and travels like a mode, that is a typical feature of the tapered fiber coupling method. As the beam goes through TIR, it is coupled to a silica micro-sphere resonator. The reflected beam at the output of the prism is confined and the WGM resonance is observed. 

An isosceles prism is diced from a z-cut stoichiometric $\mathrm{LiNbO_3}$ 0.5~mm thick wafer. This crystal features a photorefractive effect that is strong enough to obtain self-guided beam at low light power with good 2-D stability\cite{Fazio}. The optically induced waveguide were proven to last longer than a month \cite{chauvet}. In order to obtain the self-focusing and the inferred waveguides, we need a focusing nonlinearity, that can be obtained either by applying a high voltage across the sample or by slightly raising the crystal temperature to exploit the pyroelectric effect \cite{Jassem}. Note that this method was proven to be compatible with fiber coupling for infrared wavelengths\cite{Kien}, that is also an appealing feature of the fiber taper method. Since $\mathrm{LiNbO_3}$ refractive index is about n=2.2, the two equal angles of the isosceles prism are chosen to be 37.4$^\circ$ to match the refractive index of the silica microsphere. 
\begin{figure}[ht!]
\begin{center}
\includegraphics[width=1\columnwidth]{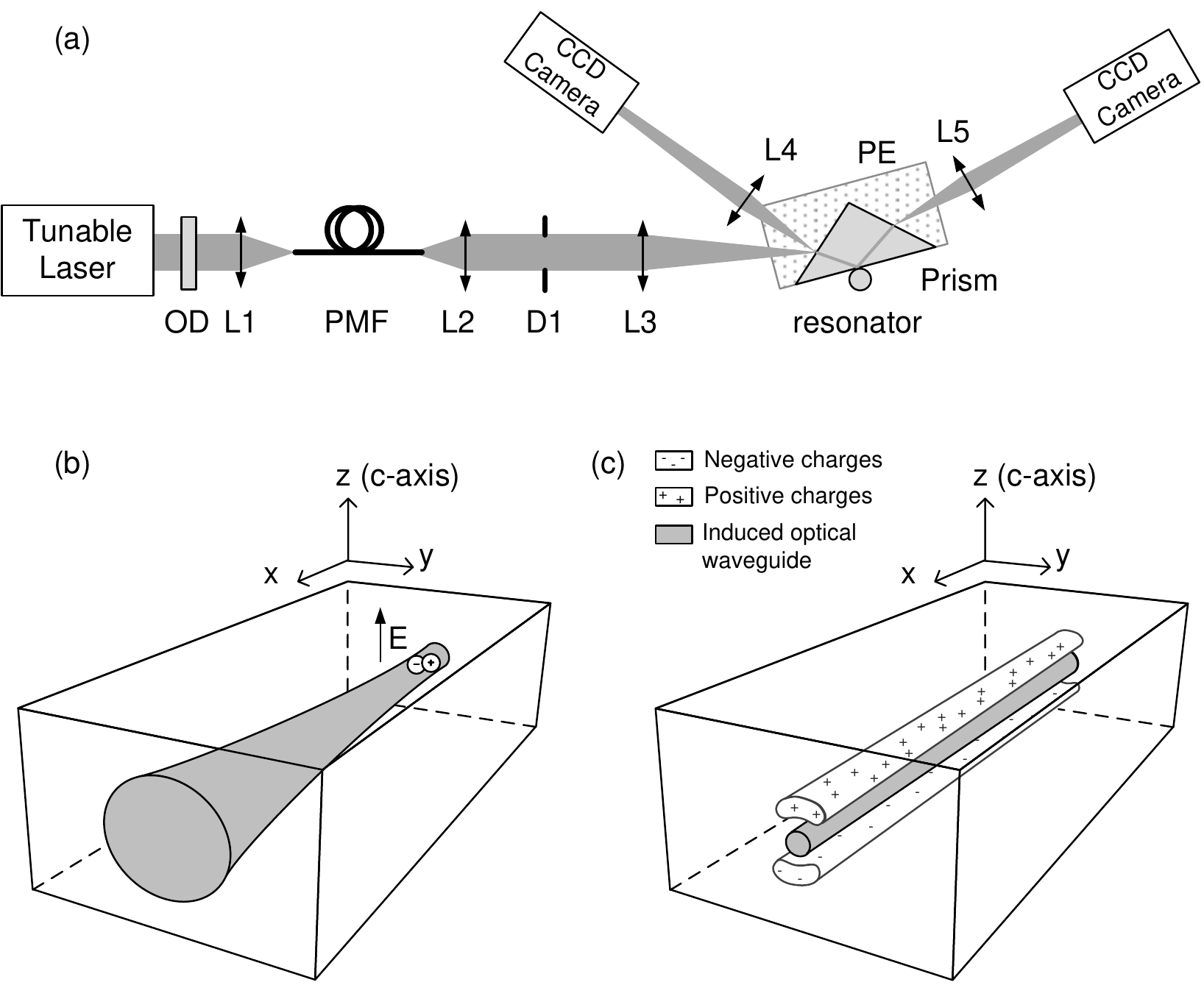}
\caption{\label{fig1} (a) Experimental setup for writing and probing waveguides inscribed by  self-trapped beams. D1,D2: Diaphragms. PC: Polarization controller. OD: Optical density. L1 ,L2, L3, L4 and L5 lenses. PBS: Polarizing Beam Splitter. HWP: Half-Wave plate. PE: Peltier Element.  (b) Diffracting beam at low power. (c) Self-trapped beam.}
\end{center}
\end{figure}
The optical set-up is described in Fig.~\ref{fig1}(a). The polarized beam of a tunable laser with a 650-660nm wavelength range (Newfocus Velocity 6305) is coupled into a Polarization Maintaining Fiber (PMF) in order to get spatial filtering and a convenient control of the polarization. The output of the fiber is imaged at the input face of the prism with two lenses (L2-L3) and a diaphragm D1 enables fine tuning of the beam waist at the prism input. Part of the beam is reflected at the prism input and is imaged through L4 on a CCD camera. The launched beam profile can thus be monitored. The transmitted light propagates inside the prism and is subject to TIR on the opposite face where the coupling with the micro-resonator occurs. The relative position between the resonator and the prism is controlled by a 3-axis stage. After reflection upon the coupling area, the reflected beam reaches the ouput face of the prism and is imaged on a second CCD thanks to lens L5. 

The optimum coupling angle is found in two steps. First we search for the proper angle that enlighten the micro-resonator at binocular sight. Then the beam is focused with L3 at the TIR point to optimize the coupling surface area, and we search for resonances. Once the proper angle is found, we pull back L3 to focus on the prism input with the desired beam width for the nonlinear regime. The self-trapped beam is obtained thanks to the photorefractive effect that will be discussed in the following paragraph. 

In the linear regime, the light diffracts freely in the crystal as shown in Fig~\ref{fig1}(b). In the non-linear regime, we slightly increase the temperature of the $\mathrm{LiNbO_3}$ crystal. A pyroelectric internal electric field $\Delta E_{py}$ appears and decreases the refractive index all over the crystal because of  the Pockels effect. However, at the input of the crystal, the focused beam generates free electrons because of the photoelectric effect. Those optical-generated free carriers drifts along the c-axis because of the pyroelectric field and recombine. The resulting space-charge field $E$ partially screens $\Delta E_{py}$. In the area where the field is screened, the refractive index is less affected which results in a localized high refractive index that eventually traps the light beam (Fig.~\ref{fig1}(c).). Because the space-charge field $E$ results from deep-center recombination, the resulting waveguide lasts even if the temperature is lowered. In our set-up, a Peltier element is placed under the crystal to control its temperature as depicted in Fig~\ref{fig1}(a). In $\mathrm{LiNbO_3}$ crystals, a moderate temperature increase $\Delta T=20^\circ$C leads to an internal electric field as high as $\Delta E_{py}=47$kV/cm \cite{Jassem,Wong}. 
\begin{figure}[ht!]
\begin{center}
\includegraphics[width=1\columnwidth]{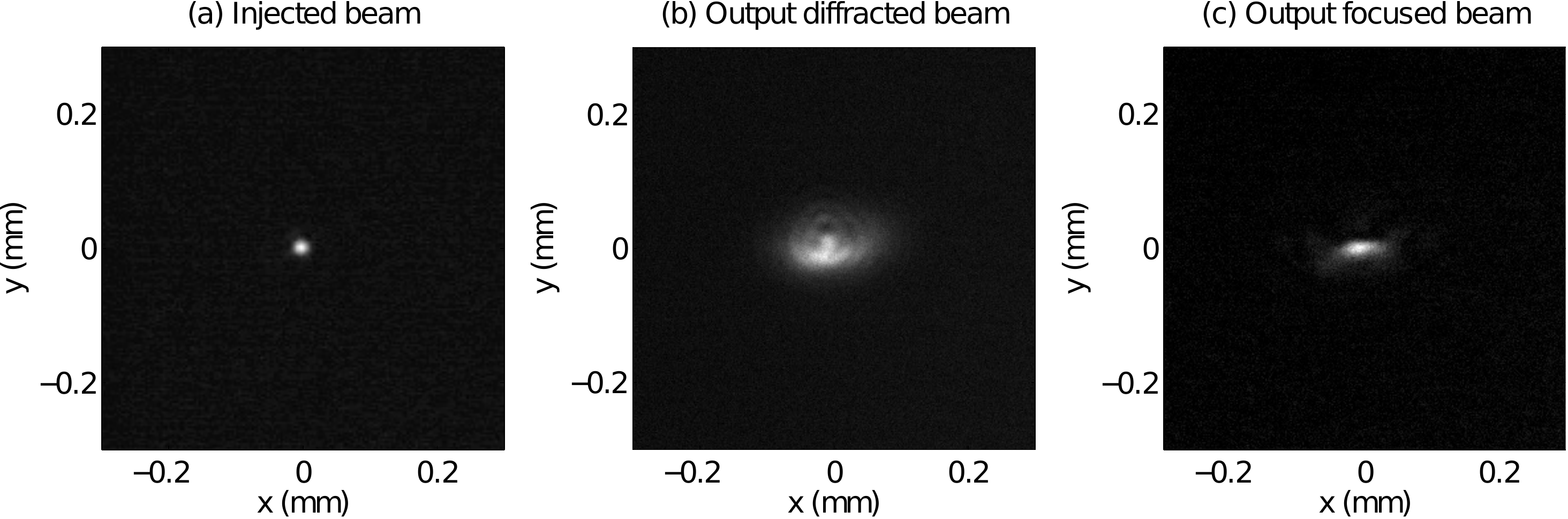}
\caption{\label{fig2} (a) Injected beam profile. (b) Diffracting beam profile at low power. (c) Self-trapped beam in nonlinear regime.}
\end{center}
\end{figure}
For strong-coupling efficiency, the beam profile should be as similar as possible to the microsphere mode size to maximize the overlap between the evanescent part of the fields. However, for WGMs, a weak-coupling regime is usually preferred to keep the Q-factor high. Analytic calculations show that high-Q WGMs are present in our laser tuning range with a beam FWHM around $5\mu$m \cite{WGMTheory}. We choose a trade-off input beam of 18~$\mu$m FWHM that is typical for self-focussing experiment in $\mathrm{LiNbO_3}$ as depicted in Fig.~\ref{fig2}(a). In Fig.~\ref{fig2}(b), the beam observed at the output face is reported in the linear regime as witnessed by the diffraction of the beam. Note that no resonance was observed in this configuration, because of the poor overlap between the diffracted beam and the WGM. 
To reach the nonlinear regime, the pyroelectric effect is triggered with a temperature increase of the $\mathrm{LiNbO_3}$ sample of $\Delta T=33^\circ$C and the light power is raised to $152~\mu$W. In about an hour, the beam is self-focused thanks to the photorefractive effect. In Fig.~\ref{fig2}(c), the crystal output beam is shown, a clear confinement is seen. It has a $16.6~\mu$m FWHM along the vertical axis (c-axis) and $42.9~\mu$m along the horizontal one. 
\begin{figure}[ht!]
\begin{center}
\includegraphics[width=1\columnwidth]{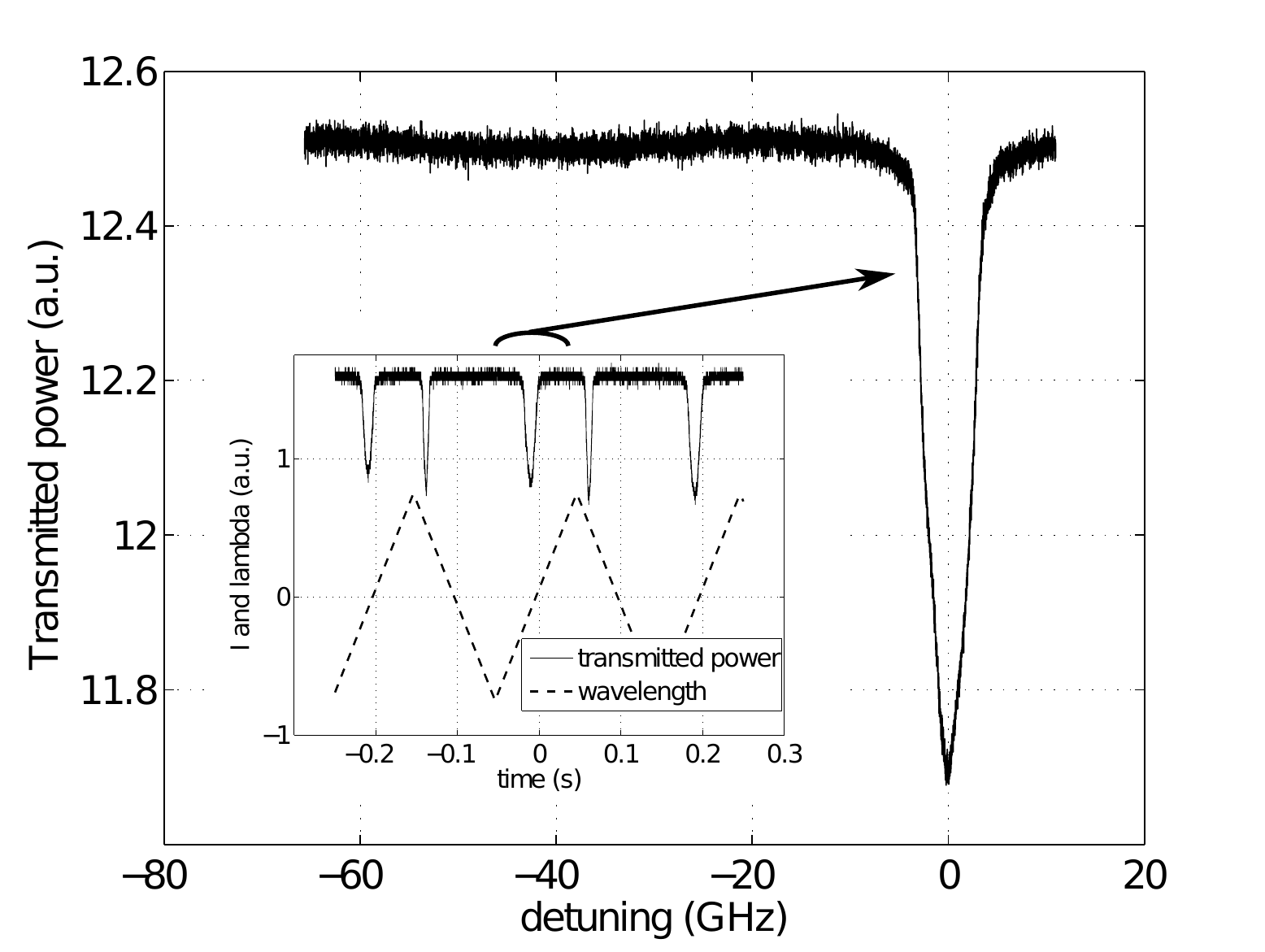}
\caption{\label{fig3} Detail of the whispering-gallery-mode resonance. Inset: Large-scale scanning of the resonance. Dashed line shows the wavelength scanning with respect to time.}
\end{center}
\end{figure}
As a result, the overlap between the beam and the WGM is improved and WGM resonances can be measured. The output camera is replaced with a photodiode and the laser frequency is periodically swept over few tens of GHz as depicted in the inset of Fig.~\ref{fig3}. The measurement is typical of a high-Q resonator hysteretic frequency response. The WGM resonance is broader for increasing frequency sweep as the thermal nonlinearity of the resonator shifts the resonance in the same direction as the laser frequency. When the laser frequency is shifted backward, the resonance is still shifted upward and does not follow the resonance anymore, as a consequence a thiner resonance peak is obtained\cite{HighQSoviet,WGMthermal}. The resonance dip reported in Fig.~\ref{fig3} corresponds to a forward sweep. A raw Q-factor of the WGM over 90000 is deduced from the 5~GHz width of the dip in transmission. An analytic calculation of the WGMs of the sphere with parameters (l,m,n) as described by Little \textit{et~al.}\cite{WGMTheory} shows that for $500<l<900$ up to 11 modes can be found within the 100~GHz scanning range of our experiment. However, only one of these modes is a fundamental mode $(l=m)$. The presence of only one dip in the measured spectrum is therefore a confirmation of the good modal selectivity of our method.

Note that once the writing phase is realized a waveguide is inscribed in the prism. This waveguide has the proper characteristics to guide light with the right trajectory and the adapted mode profile to couple efficiently to the resonator. It takes advantage of the unique properties of self-alignment and confinement provided by trapped beams.

To conclude, we have reported light coupling in a WGM resonator with the help of a self-trapped beam. This original configuration gathers both the flexibility of the prism coupling and the benefit from the waveguide confinement. In addition, it is mechanically more robust than the fiber-taper method. Improvements of the method are foreseen by optimizing the output beam shape through a better control of the dynamic of the nonlinear effect. The concept can be extended to other photosensitive materials such as photopolymers to fabricate permanent sturdy coupling to resonators.



We are grateful for financial support by Agence National de la Recherche for
project ORA (ANR 2010 BLAN-0312). This work was realized in the framework of the
French Labex ÒActionÓ and was partly supported by the
RENATECH network and its FEMTO-ST technological facility.
Kien Phan Huy thanks Thibaut Sylvestre for his technical help on the injection.


\begin{thebibliography}{99}

\bibitem{CouplagePrism} P. K. Tien and R. Ulrich,  J. Opt. Soc. Am., \textbf{60}(10), 1325--1337 (1970).

\bibitem{HighQSoviet} V. B. Braginsky, M. L. Gorodetsky and V.S Ilchenko, Phys. Lett. A, \textbf{137}(7), 393--397 (1989).

\bibitem{TaperPolished}  G. Griffel, S. Arnold, D. Taskent, A. Serpeng\"uzel, J. Connolly and N. Morris,  Opt. Lett., \textbf{21}(10), 695--697(1996).

\bibitem{TaperKnight} J. C. Knight, G. Cheung, F. Jacques, and T. A. Birks,  Opt. Lett., \textbf{20}(15), 1129--1131(1997).

\bibitem{Fazio} E. Fazio, F. Renzi, R. Rinaldi, M. Bertolotti, M. Chauvet, W. Ramadan, A. Petris, and V. I. Vlad,  Appl. Phys. Lett. \textbf{85}, 2193--2195 (2004).

\bibitem{chauvet} M. Chauvet,  J. Opt. Soc. Am. B, \textbf{20}, 2515--2522 (2003).

\bibitem{Jassem} J. Safioui, F. Devaux, and M. Chauvet, Opt. Express \textbf{17}, 22209--22216 (2009).

\bibitem{Kien} K. Phan~Huy, J. Safioui, B. Guichardaz, F. Devaux, and M. Chauvet,  Appl. Opt. \textbf{51}, 4353--4358 (2012).

\bibitem{Wong} K. K. Wong, ``Properties of Lithium Niobate'', (Academic, 2002)

\bibitem{WGMTheory} B. E. Little, J.-P. Laine, and H. Haus, J. Lightw. Technol.  \textbf{17}(4), 704--715, (1999).

\bibitem{WGMthermal} T. Carmon, L. Yang, and K. J. Vahala, Opt. Express  \textbf{12}(20), 4742--4750, (2004).


\end{thebibliography}

\newpage



\begin{thebibliography}{99}

\bibitem{CouplagePrism} P. K. Tien and R. Ulrich, ``Theory of Prism-Film Coupler and Thin Film Light Guides,'' J. Opt. Soc. Am., \textbf{60}(10), 1325--1337 (1970).

\bibitem{HighQSoviet} V. B. Braginsky, M. L. Gorodetsky and V.S Ilchenko, ``Quality-factor and nonlinear properties of optical whispering-gallery modes,'' Phys. Lett. A, \textbf{137}(7), 393--397 (1989).

\bibitem{TaperPolished}  G. Griffel, S. Arnold, D. Taskent, A. Serpeng\"uzel, J. Connolly and N. Morris, ``Morphology-dependent resonances of a microsphereÐoptical fiber system,'' Opt. Lett., \textbf{21}(10), 695--697(1996).

\bibitem{TaperKnight} J. C. Knight, G. Cheung, F. Jacques, and T. A. Birks, ``Phase-matched excitation of whispering-gallery-mode
resonances by a fiber taper,'' Opt. Lett., \textbf{20}(15), 1129--1131(1997).


\bibitem{Fazio} E. Fazio, F. Renzi, R. Rinaldi, M. Bertolotti, M. Chauvet, W. Ramadan, A. Petris, and V. I. Vlad, ``Screening photovoltaic bright solitons in lithium niobate and associated
single-mode waveguides,'' Appl. Phys. Lett. \textbf{85}, 2193--2195 (2004).

\bibitem{chauvet} M. Chauvet, ``Temporal analysis of open-circuit dark photovoltaic spatial solitons,'' J. Opt. Soc. Am. B, \textbf{20}, 2515--2522 (2003).

\bibitem{Jassem} J. Safioui, F. Devaux, and M. Chauvet, ``Pyroliton: pyroelectric
spatial soliton,'' Opt. Express \textbf{17}, 22209--22216 (2009).

\bibitem{Kien} K. Phan~Huy, J. Safioui, B. Guichardaz, F. Devaux, and M. Chauvet, ``Writing and probing light-induced waveguides thanks to an endlessly single-mode photonic crystal fiber,'' Appl. Opt. \textbf{51}, 4353--4358 (2012).



\bibitem{Wong} K. K. Wong, ``Properties of Lithium Niobate'', (The Institution of Engineering and Technology, 2002)

\bibitem{WGMTheory} B. E. Little, J.-P. Laine, and H. Haus, ``Analytic theory of coupling from tapered fibers and half-blocks into microsphere resonators,'' J. Lightw. Technol.  \textbf{17}(4), 704--715, (1999).

\bibitem{WGMthermal} T. Carmon, L. Yang, and K. J. Vahala, ``Dynamical thermal behavior and thermal self stability
of microcavities,'' Opt. Express  \textbf{12}(20), 4742--4750, (2004).


\end{thebibliography}
\end{document}